\documentclass{article} 
\usepackage{epsfig}

\usepackage{amssymb}
 \usepackage{graphicx}
\usepackage{amsmath}
 
\usepackage{amsmath}
\usepackage{amsfonts} 

\usepackage{a4}
\newcommand{\beq}{\begin{equation}}
\newcommand{\eeq}{\end{equation}}
\newcommand{\beqs}{\begin{eqnarray}}
\newcommand{\eeqs}{\end{eqnarray}}

\newcommand{\insertplot}[5]{\begin{figure}
 \hfill\hbox to 0.05in{\vbox to #5in{\vfill
 \inputplot{#1}{#4}{#5}}\hfill}
 \hfill\vspace{-.1in}
 \caption{#2}\label{#3}
 \end{figure}}
 \newcommand{\inputplot}[3]{
 \special{ps: plotfile #1}
}
\newcounter{fig}

\renewcommand{\t}{\theta}

\newcommand{\ee}{\end{equation}}
\newcommand{\eea}{\end{eqnarray}}
\newcommand{\be}{\begin{equation}}
\newcommand{\bea}{\begin{eqnarray}}

\begin{document}

\title{ 
Axially symmetric static scalar solitons \\ and black holes with scalar hair
}

\author{
{\large  Burkhard Kleihaus,} 
{\large  Jutta Kunz,} 
{\large Eugen Radu} 
and {\large Bintoro Subagyo}  
\\ 
 {\small  Institut f\"ur Physik, Universit\"at Oldenburg, Postfach 2503
D-26111 Oldenburg, Germany} 
}

\maketitle

\begin{abstract} 
We construct static, asymptotically flat
black hole solutions with scalar hair. 
They evade the no-hair theorems
by having a scalar potential which is not strictly positive.
By including an azimuthal winding number in the scalar field  ansatz, we find
hairy black hole solutions
which are static but axially symmetric only.
These solutions possess a globally regular limit, describing scalar solitons. 
A branch of axially symmetric black holes is found to possess a 
positive specific heat.
\end{abstract}

 \medskip

{\bf Introduction.--}
The energy conditions are an important  ingredient of various significant
results in general relativity \cite{Hawking:1973uf}.
Essentially, they imply that some linear combinations of the energy-momentum
tensor of the matter fields should be positive, or at least non-negative.
However, over the last decades, it has become 
increasingly obvious that these conditions can be violated,
even at the classical level.
Remarkably enough, the violation may occur also for the simplest case 
of a scalar field 
(see $e.g.$ \cite{Barcelo:2002bv} for a discussion of these aspects).

Once we give up the energy conditions (and in particular the weak one), 
a number of results in the literature
show that the asymptotically flat black holes may 
possess scalar hair\footnote{One has to remark that the existence of black holes with 
scalar hair is perhaps the mildest consequence of
giving up the energy conditions, see $e.g.$ the discussion in 
\cite{Hertog:2003ru}.}, which  otherwise is forbidden by a number of
well-known theorems \cite{Bekenstein:1996pn}.
Restricting to the simplest case of a minimally coupled scalar field
with a scalar potential which is not strictly positive,
this includes both analytical 
\cite{Bechmann:1995sa}, 
\cite{Dennhardt:1996cz},  
\cite{Nikonov:2008zz},
\cite{Bronnikov:2001ah},
\cite{Anabalon:2012ih}
and numerical 
\cite{Nucamendi:1995ex},
\cite{Corichi:2005pa}
 results.

Interestingly, in the limit of zero event horizon radius, 
some of these hairy black holes  describe globally regular, particle-like objects,
the so-called '{\it scalarons}' \cite{Nucamendi:1995ex}.
At the same time, a complex scalar field is known for long time 
to possess non-topological solitonic solutions \cite{Lee:1991ax},
even in the absence of gravity.
These are the Q-balls introduced by Coleman in \cite{Coleman:1985ki}.
Such configuration owe their existence
to a  harmonic time dependence of the scalar field and 
possess a positive energy density.

However, as argued below, the Q-balls can be reinterpreted as non-gravitating scalarons.
The scalar field is static in this case 
and has a potential which  takes negative values as well.  
As expected, the scalarons possess gravitating generalizations.
However, different from the standard Q-ball case \cite{Pena:1997cy}, their regular origin 
can be replaced with an event horizon.
In this work we study such solutions for 
the simple case of a  massive complex scalar field
with a  negative  quartic self-interaction term in the potential.
Apart from spherically symmetric 
configurations,
we construct solitons and hairy black hole solutions
which are static but axially symmetric only.

{\bf The model.--}
Let us consider the action of a self-interacting complex scalar field 
$\Phi$ coupled to Einstein gravity in four spacetime dimensions,
\begin{equation}
\label{action}
S=\int  d^4x \sqrt{-g}\left[ \frac{1}{16\pi G}R
   -\frac{1}{2} g^{\mu\nu}\left( \Phi_{, \, \mu}^* \Phi_{, \, \nu} + \Phi _
{, \, \nu}^* \Phi _{, \, \mu} \right) - U
 \right] , 
\end{equation}
where $R$ is the curvature scalar,
$G$ is Newton's constant and
the asterisk denotes complex conjugation.
Using the principle of variation,
one finds the coupled Einstein--Klein-Gordon equations
\begin{equation}
\label{Einstein-eqs}
E_{\mu\nu}=R_{\mu\nu}-\frac{1}{2}g_{\mu\nu}R- 8 \pi G~T_{\mu\nu}=0,
~~\frac{1}{\sqrt{-g}} \partial_\mu \big(\sqrt{-g} \partial^\mu\Phi \big)=\frac{\partial U}{\partial\left|\Phi\right|^2} \Phi,
\end{equation}  
where $T_{\mu\nu}$ is the
 stress-energy tensor  of the scalar field
\begin{eqnarray}
\label{tmunu} 
T_{\mu \nu}  
=
\left(
 \Phi_{, \, \mu}^*\Phi_{, \, \nu}
+\Phi_{, \, \nu}^*\Phi_{, \, \mu} 
\right )
-g_{\mu\nu} \left[ \frac{1}{2} g^{\alpha\beta} 
\left( \Phi_{, \, \alpha}^*\Phi_{, \, \beta}+
\Phi_{, \, \beta}^*\Phi_{, \, \alpha} \right)+U
\right]
 \ .
\end{eqnarray}
In the above relations $U$ denotes the scalar field potential,
which, in order to retain the $U(1)$ symmetry of the
whole Lagrangian,   must be a function of $|\Phi|^2$.
In what follows, we assume that $U$ can be written as
\begin{eqnarray}
\label{pot}
U=\sum_{k\geq 1} c_k |\Phi|^{2k},
\end{eqnarray}
the $k>1$ terms taking effectively into account various interactions.
Of interest here is the case of a potential
which is not strictly positive definite.
Then  the polynomial $F(x)=\sum_{k\geq 1} c_k x^k$ is
negative for some range of $x>0$, which implies that at least one of coefficients $c_k$ is smaller than zero.
Since we assume\footnote{This can always be realized $via$ a redefinition of the scalar field.} $\Phi\to 0$ asympotically,
the requirement to obtain a bound state 
imposes $c_1=\mu^2 > 0$, with $\mu$ the scalar field mass.

{\bf Flat space solitons: Q-balls as scalarons.--}
Let us start our discussion with the simple observation that 
when ignoring the gravity effects,
a class of solutions of
the model (\ref{action})  
is already known.
We recall that
in a flat spacetime background, the Klein-Gordon equation possesses
non-topological soliton solutions--the so-called Q-balls, in which case 
the scalar has a harmonic time dependence, $\Phi=\phi( x) e^{-i wt}$
\cite{Coleman:1985ki}  (with $x^\mu=(x^a,t)$).
As a result, the solutions possess a nonvanishing conserved Noether charge,
$Q=2 w \int d^3  x |\phi|^2$. 
Then, even though $\Phi$ depends on time, the energy-momentum tensor 
$T_\mu^\nu$ is time independent and
 the effective action of this model reads
 \begin{equation}
\label{action1}
S_Q=-\int  d^3  x dt \left[ \phi_{, \, a}^* \phi^{, \, a}-w^2 |\phi|^2+ U
 \right] , 
\end{equation}
.
The Q-balls have been extensively discussed
in the literature 
(see the review work \cite{Lee:1991ax}, \cite{Radu:2008pp}) and they
have found a variety of physically interesting applications\footnote{For example, the $Q-$ball solutions appear in supersymmetric 
generalizations of the standard model 
\cite{Kusenko:1997zq}.
Also, they may be responsible for the generation of
baryon number or may even be regarded as candidates for dark matter  
\cite{Kusenko:1997si}.}.
If one assumes a potential of the form (\ref{pot}), then 
  $U$ necessarily contains powers of $|\Phi|^2$ higher than two,
 the usual choice in the literature being
$
U=\mu^2 |\Phi|^2  -\lambda |\Phi|^{4}+\nu  |\Phi|^{6},
$
 with $\lambda>0,~\nu>0$ and $\lambda^2< 4 \mu^2 \nu$ for a positive potential.
 
However, one can see from (\ref{action1}) that $w^2$ acts as an effective tachyonic contribution to the  mass term,
and thus it can be absorbed into $\mu^2$.
The scalar field is static  in this case,
$\Phi=\phi(x^a)$ and thus the Noether charge vanishes.
Therefore all Q-ball solutions in a flat spacetime background can be interpreted as static 
scalar solitons, $i.e.$ they become scalarons 
in a model with a shifted scalar field mass, for a new potential $U=U_{(Q-ball)}-w^2 |\phi|^2$.
Note that although $\phi$ satisfies the same equation as before, the energy-momentum tensor and 
the total mass of the scalarons are different. 
Also, as implied by the Derick-type virial identity
  \begin{eqnarray}
 \label{vir-flat}
\int d^3 x \left[ \phi_{, \, a}^* \phi^{, \, a} + 3U
 \right] =0, 
\end{eqnarray}
the redefined potential $U$ is necessarily negative for 
some range of $|\phi|^2$ which is realised by the solutions. 
However, the scalarons' total mass is strictly positive,
  \begin{eqnarray}
 \label{mass-flat}
M=\int d^3  x \left[ \phi_{, \, a}^* \phi^{, \, a} + U
 \right] = \frac{2}{3} \int d^3  x~ \phi_{, \, a}^* \phi^{, \, a}.
\end{eqnarray}
Finally, we mention also that the mass of the scalarons is fixed by the mass $M_Q$ and the Noether charge $Q$
of the Q-balls, $M=M_{Q}-w Q$.
 
{\bf Spherically symmetric, gravitating solutions.--}
However, the curved spacetime  scalarons cannot be interpreted as boson stars and 
thus require
a separate study.
For example,   
following \cite{Pena:1997cy}, one can show that, even in the absence of backreaction,
 one
cannot add a black hole horizon inside a Q-ball\footnote{Note, however, the boson shells harbouring black  holes in \cite{Kleihaus:2009kr}.
These solutions require  a $V$-shaped scalar potential which is not of the form (\ref{pot}). } (this
follows essentially because a Q-ball possesses a $e^{-i wt}$ time dependence and $t\to \infty$ at the horizon
of a black hole).
However, this obstruction does not apply to scalarons, 
which possess finite energy, regular generalizations
also for a static black hole background. 

Let us start with a discussion of the spherically symmetric gravitating solutions 
of the 
model (\ref{action}). 
These configurations are easier to study and
some of their properties seem to be generic.
A sufficiently general metric ansatz in this case reads
 \begin{eqnarray}
 \label{sph0}
 ds^2= g_{rr}{dr^2} +g_{\Omega\Omega}  d\Omega_2^2+g_{tt}dt^2, 
  \end{eqnarray}
(with $d\Omega_2^2=d\theta^2+\sin^2 \theta d\varphi^2$), and 
the scalar field is a function of $r$ only, $\Phi=Z(r)$.
One possible direction here is to choose a metric gauge with
$-g_{tt}=1/g_{rr}=V(r)$, 
$g_{\Omega\Omega} =P^2(r)$.
 Then the Einstein equations imply the relation
$\frac{P''}{P}+8\pi G Z'^2 =0$
(where the prime denotes a derivative with respect to $r$).
The approach taken in 
\cite{Bechmann:1995sa}, 
\cite{Dennhardt:1996cz} (see also \cite{Nikonov:2008zz}) 
is to postulate an expression for the scalar field
and to use this relation to derive $P$.
In the next step, 
the remaining Einstein equations
are used to reconstruct the
scalar potential $U$
and the metric function $V$ compatible with $Z$ and $P$.
This approach has the advantage to lead to partially
closed form solutions, but the resulting 
expressions are very complicated; also the potential cannot be written in the form (\ref{pot}).

In what follows we solve the field equations numerically for a given potential.
In this case it is convenient to work in Schwarzschild-like coordinates with 
  \begin{eqnarray}
 \label{sph1}
 g_{rr}=\frac{1}{N(r)},~~g_{\Omega\Omega}=r^2,~~g_{tt}=-N(r)\sigma^2(r),~~{\rm with}~~N(r)=1-\frac{2m(r)}{r}, 
  \end{eqnarray}
  where $m(r)$ may be interpreted as the total
mass-energy within the radius $r$; its derivative $m'$ is proportional to the energy density $\rho=-T_t^t$.
 Then the field equations (\ref{Einstein-eqs}) reduce to
\begin{eqnarray}
\label{sph3}
&&
m'=4 \pi G r^2(N Z'^2+U ),
~~~
\sigma'=8 \pi G r\sigma Z'^2,~~
Z''+(\frac{\sigma'}{\sigma}+\frac{N'}{N}+\frac{2}{r})Z'
-\frac{1}{N}\frac{\partial U}{\partial Z^2}Z=0.
\end{eqnarray} 
For a generic $U$, it is possible to write an approximate form of the solutions close to the horizon
(or at the origin) and also for large $r$.
These asymptotics are connected by constructing numerically
the solutions, which requires to specify the expression for the scalar field potential.

The horizon of the black holes is located at $r=r_H>0$, where the solutions have a power-series expansion
\begin{eqnarray} 
\label{sol-hor}
m(r)=\frac{r_H}{2}+m_1(r-r_H)+\dots,~~\sigma(r)=\sigma_0 +8\pi G\sigma_0 r_H z_1^2(r-r_H)+\dots,~~
Z(r)=z_0+z_1(r-r_H)+\dots,
\end{eqnarray}
 in terms of two arbitrary parameters $Z(r_H)=z_0$ and $\sigma(r_H)=\sigma_0$ (with
$
m_1=4 \pi G r_H^2 U(z_0),$ and $z_1=\frac{r_H}{1-2m_1}\frac{\partial U}{\partial Z^2}\big|_{z_0}z_0).
$
One can write an approximate form of the solutions also for $r\to \infty$, with
\begin{eqnarray} 
\label{sol-inf}
m(r)=G M -4 \pi G \mu \bar z_1^2 e^{-2\mu r}+\dots,
\log \sigma(r)= -8\pi G \bar z_1^2 \mu 
\left(\frac{ e^{-2\mu r}}{r }+\mu Ei(-2\mu r) \right)+\dots,
Z(r)=\bar z_1 \frac{e^{-\mu r}}{r}+\dots,
\end{eqnarray}
with $Ei(x)$ the exponential integral function \cite{grad}; 
$M$, $\bar z_1$ are two parameters fixed by the numerical calculations,
$M$ corresponding to the total mass of the solutions.

The Hawking temperature and event horizon area of a spherically symmetric black hole are
 \begin{eqnarray} 
\label{q1}
T_H=\frac{\sigma'(r_H)}{4\pi r_H}(1-2m'(r_H)),~~A_H=4\pi r_H^2,
\end{eqnarray}
 the entropy of the solutions being $S=A_H/4G$.

 \setlength{\unitlength}{1cm}
\begin{picture}(8,6) 
\put(-0.5,0.0){\epsfig{file=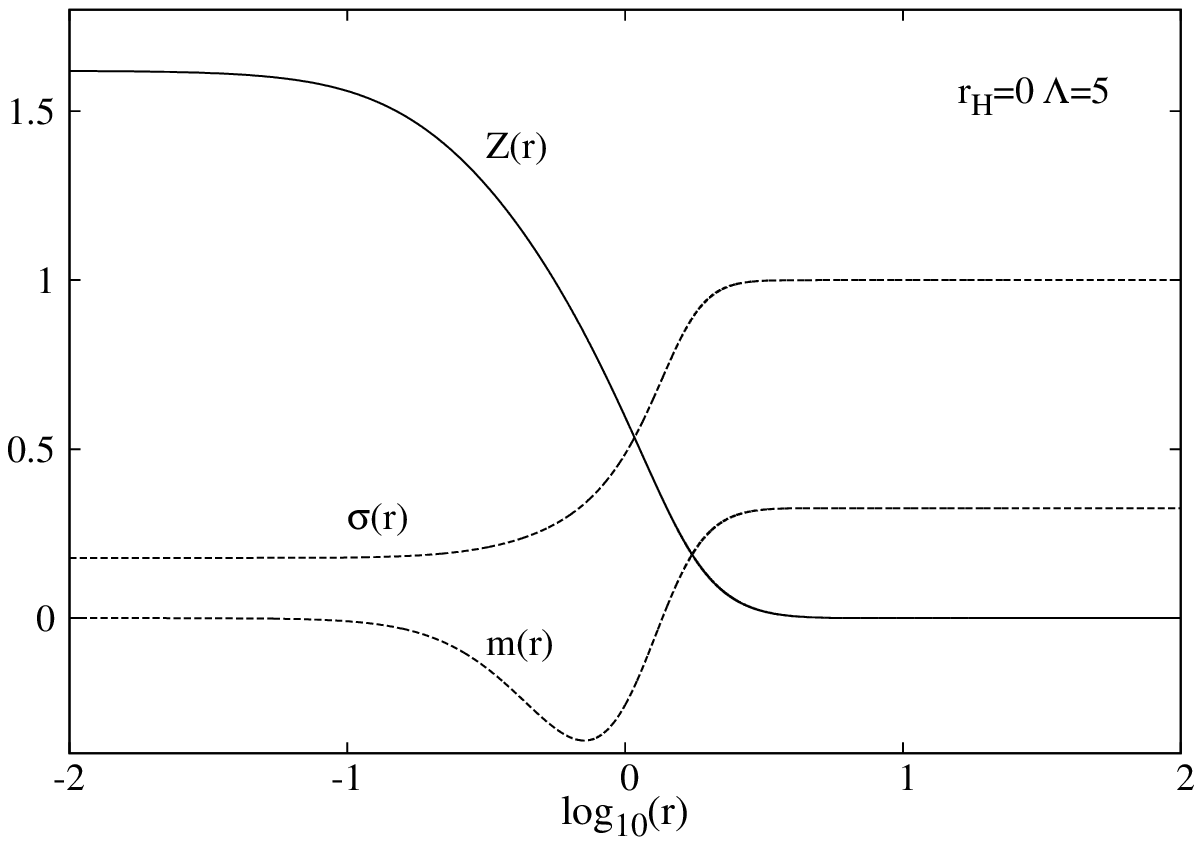,width=8cm}}
\put(8.1,0.0){\epsfig{file=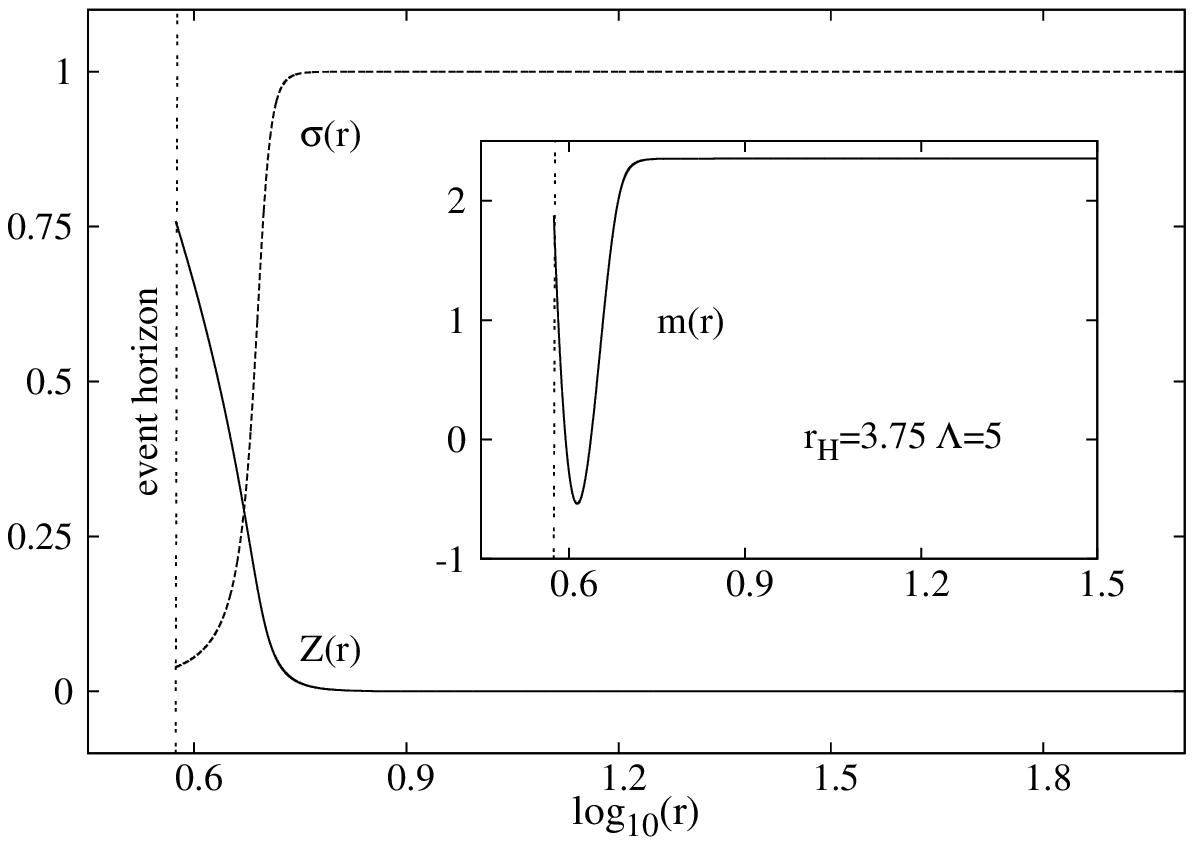,width=8cm}}
\end{picture}
\\
{\small {\bf Figure 1.} 
The profiles of a spherically symmetric scalar soliton (left) and  a hairy black hole solution (right)  
 versus the radial coordinate.  
}
\vspace{0.5cm}
%
\\
For any symmetry, the static  black hole solutions 
satisfy the Smarr relation \cite{Bardeen:1973gs}
\begin{eqnarray}
\label{smarr}
M=2T_H S+M^{(ext)} ,
\end{eqnarray} 
 where
 \begin{eqnarray}
 M^{(ext)} =- \int_{r>r_H} d^3 x \sqrt{-g}(2T_t^t-T_\mu^\mu),
\end{eqnarray} 
is the contribution to the
total mass of the matter outside the event horizon
(with $\int d^3 x=4\pi \int_{r_H}^\infty dr$ for spherically symmetric configurations),
and the first law of thermodynamics \cite{Heusler:1993cj},
 \begin{eqnarray}
 \label{fl}
dM=T_H dS.
\end{eqnarray} 
Also, 
by using the approach in \cite{Heusler:1996ft},
one can prove the following virial identity:
\begin{eqnarray}
\label{vir1}
\int_{r_H}^\infty dr~
\sigma r^2
\bigg [
Z'^2\left(1-\frac{r_H}{r}(1+N) \right)+U (3-\frac{2r_H}{r})
\bigg ]
=0,
\end{eqnarray} 
 ($r_H=0$ gives the corresponding relation for the solitonic case).

For a quantitative study of the solutions, we need to specify the expression
for $U$. 
The results reported in this work correspond 
to the simplest potential allowing for $U<0$, with 
\begin{eqnarray}
\label{pot1}
U = \mu^2 |\Phi|^2 -\lambda |\Phi|^4,
\end{eqnarray}
where $\lambda$ is a strictly positive parameter.
The existence of hairy black hole
solutions for this choice of the potential has been noticed in \cite{Gubser:2005ih};
a possible physical justification for this expression of $U$ can also be found there.
 
In this case, the system possesses two scaling symmetries (these symmetries
are independent of any specific ansatz and hold also for the axially symmetric solutions below):
 \begin{eqnarray}
 \label{s1} 
(i)~~  x^a \to  x^a c,~~\mu\to \mu/c,~\lambda\to \lambda/c,~~ 
 {\rm and}~~~
 (ii)~~\Phi\to \Phi c,~~\lambda \to \lambda/c^2,~~G\to G/c^2 ~,
 \end{eqnarray} 
(also with $m \to  m c$ for (i); note the invariant parameters are not shown here)
 which are used to define a dimensionless radial variable $r\to r  \mu$ and 
 a scaled scalar field,  
 $\Phi \to \Phi \sqrt{4\pi}/ M_{Pl}$ 
(with $M_{Pl}=1/\sqrt{G}$  the Planck mass for the units employed in this work).
 Then, for a given $r_H$, families of solutions can be parametrized by the single
  dimensionless quantity
 \begin{eqnarray}
\label{alfa}
\Lambda=\frac{\lambda M_{Pl}^2}{4\pi \mu^2}.
\end{eqnarray} 

\newpage
 \setlength{\unitlength}{1cm}
\begin{picture}(8,6) 
\put(-0.5,0.0){\epsfig{file=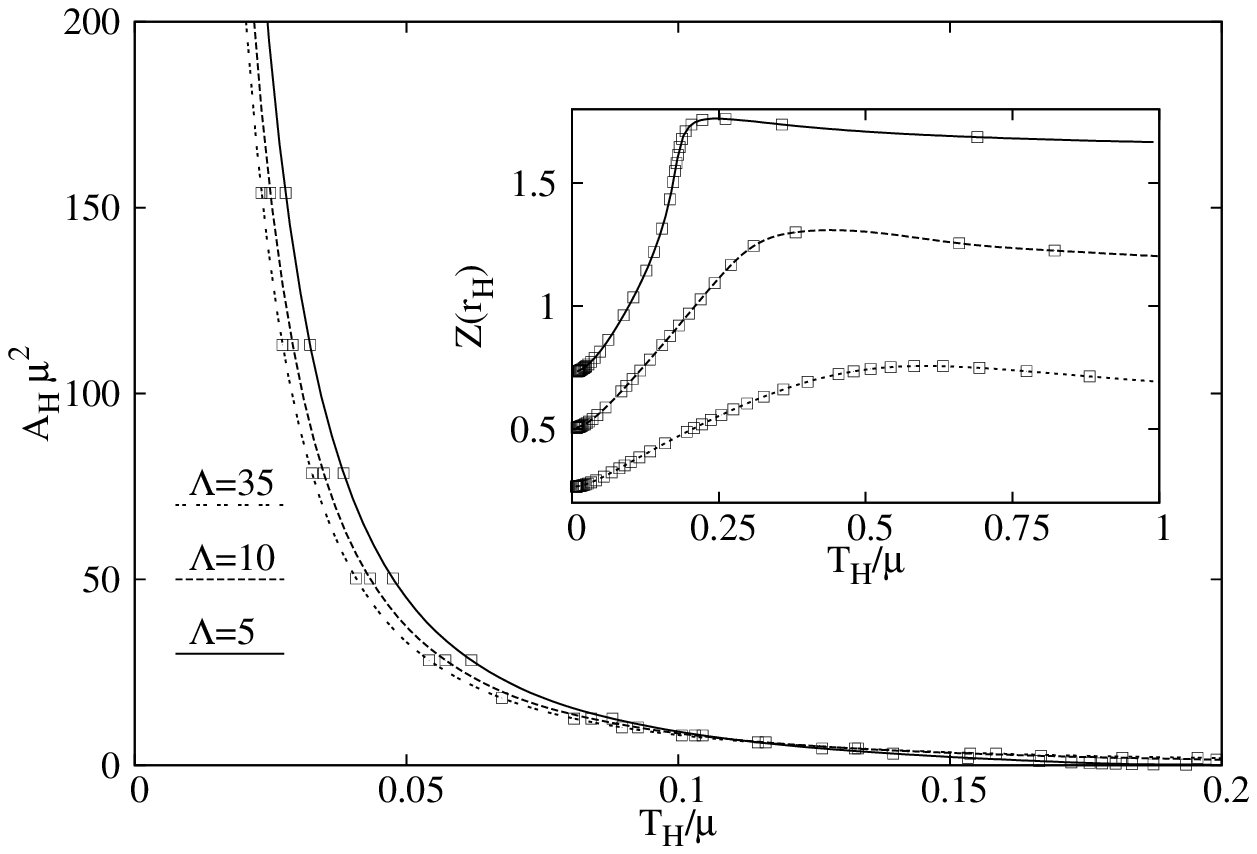,width=8cm}}
\put(8.1,0.0){\epsfig{file=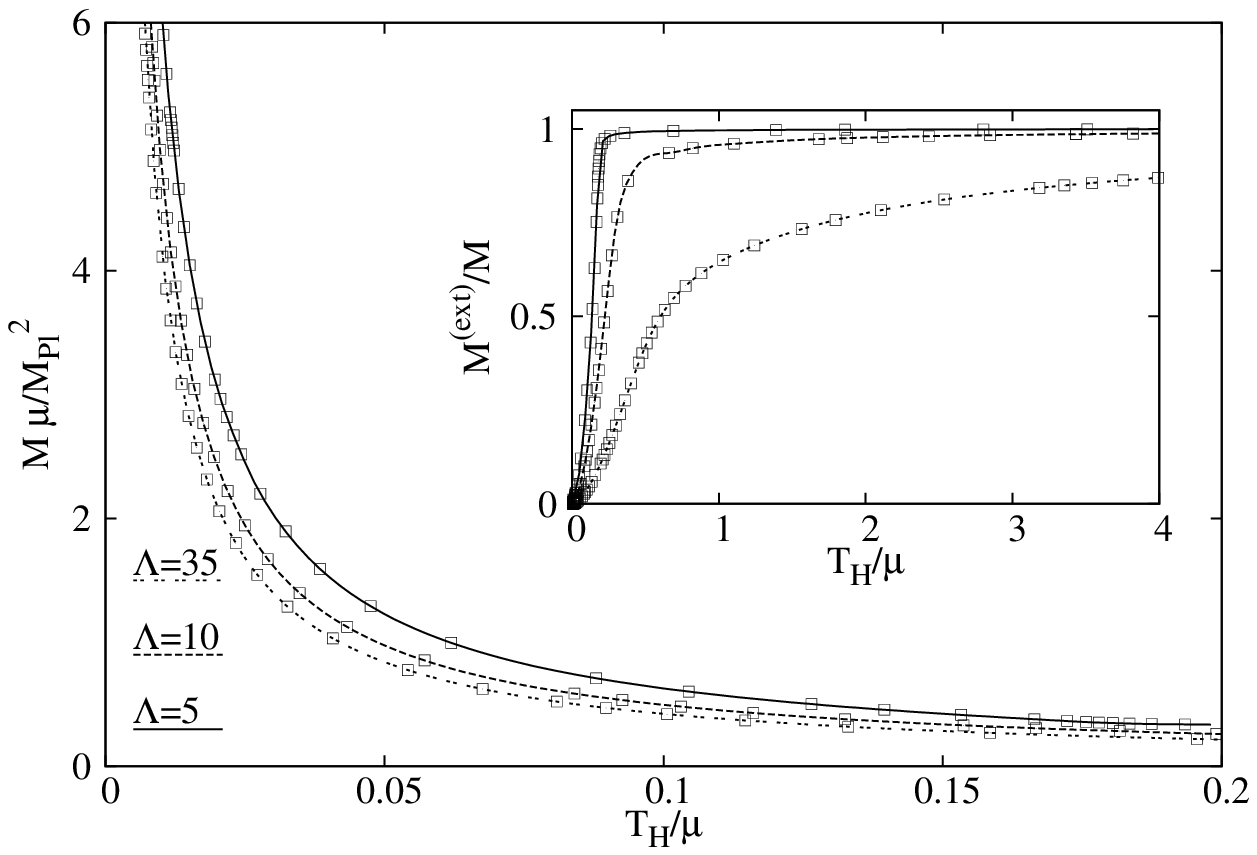,width=8cm}}
\end{picture}
\\
{\small {\bf Figure 2.} 
Some properties of
spherically symmetric black holes with scalar hair are shown as functions of the Hawking temperature
for several values of the  dimensionless self-coupling constant $\Lambda$.
}
\vspace{0.5cm}
\\
Also, all quantities in this work are given in natural units set by   $\mu$ and $G$.
   
The properties of the spherically symmetric solutions can be
summarized as follows\footnote{The solutions reported in this work have a nodeless scalar field. However,
solutions with $Z$ taking both positive and negative values do also exist, 
but we did not attempt to study them systematically.}.
First, the model possesses soliton configurations ($r_H=0$), which are the gravitating
scalarons.
These  globally regular configurations possess  at $r=0$
a power-series expansion with (here we employ dimensionless variables)
$m(r)=\frac{1}{3} b^2(1-b^2\Lambda)r^3+ \mathcal O(r^5),~~\sigma(r)=\sigma_0+2 \sigma_0z_2^2 r^4+  \mathcal O(r^6),~~
 Z(r)=b+z_2 r^2+ \mathcal O(r^4)$ (with $z_2=b(1-2b^2\Lambda)/6$), while for large $r$ the expressions (\ref{sol-inf}) are still valid.

The only input parameter in this case is $\Lambda$.
Rather unexpected, we have found that 
the scalarons do not exist for an arbitrarily small coefficient of the quartic term in the potential. 
That is, we could find solutions with the right asymptotics  
 for $\Lambda > \Lambda_{min}\simeq 1.63$ only.
The energy  of scalarons is localized  in a small region around the origin,
both the scalar field and the metric functions possessing a  nontrivial dependence on $r$ (with $\sigma(0)\neq 1)$.
The profile of a typical scalaron is shown in Figure 1 (left).
One can see that the solution violates the weak energy condition, 
since $m' \sim \rho<0$ for some
range of $r$.
However, we have found that the total mass $M$ is always positive, which is consistent with the general
results \cite{Hertog:2003ru}.

Similar to the case of other static 
solitons \cite{Volkov:1998cc}, for any scalaron with a given $\Lambda$,
one can replace the regular origin with a black hole.
As expected from above, the solutions exist only for values of $\Lambda$
greater than a minimal value  which depends on $r_H$.
The profile of a typical black hole is shown in Figure 1 (right).

The control parameter in this case is $r_H$,
the event horizon radius, with $A_H\to 0$ and $T_H\to \infty$
as $r_H\to 0$.
Solutions are likely to exist for an arbitrarily large event horizon radius,
although the numerical calculations become difficult for large $r_H$.
As $r_H$ increases, both the mass and event horizon area increase,
while the temperature decreases (see Figure 2).
Thus, as expected, these solutions possess a negative specific heat.
 Furthermore, it turns out that
the free energy $F=M-T_H S$  
of a Schwarzschild solution  is larger than the free energy
of a hairy solution with the same temperature.
Then, these are likely to be unstable and to decay to the Schwarzschild solution.

Indeed, we have found that the spherically symmetric solutions are unstable
against linear
fluctuations. In examining time-dependent fluctuations around the solutions discussed above,
all field variables are written as the sum of the static equilibrium solution whose stability we
are investigating and a time dependent perturbation. By following the standard methods,
we derive linearized equations for $\delta \sigma(r, t)$, $\delta N(r, t)$ and $\delta Z(r, t)$.
These equations
imply that both $\delta \sigma(r, t)$ and $\delta N(r, t)$ are determined by $\delta Z(r, t)$. 
For a harmonic time dependence
$e^{-i \Omega t}$, the linearized scalar field equation reduces to a standard Schr\"odinger equation
 \begin{eqnarray}
\label{Seq}
\left \{-\frac{d^2}{d\rho^2} +V_{{\rm eff}} \right \}\Psi(\rho)=\Omega^2 \Psi(\rho) ,
\end{eqnarray}
where  $\Psi = \delta Z e^{i\Omega t}/r $  and a new  radial coordinate is introduced, $d/d\rho = N\sigma d/dr$.
The effective potential in (\ref{Seq}) is given by
$
V_{{\rm eff}}=\sigma^2 N 
\left[
\frac{N}{r}(\frac{N'}{N}+\frac{\sigma'}{\sigma})-4 r N (\frac{N'}{N}+\frac{\sigma'}{\sigma}+1)Z'^2
+8r Z'\frac{\partial U}{\partial Z}+\frac{1}{2}\frac{\partial^2 U}{\partial Z^2} 
\right ].
$
By solving numerically the above  Schr\"odinger equation with suitable boundary conditions (namely $\Psi(r_H)=0$
and $\Psi(r) \to 0$ as $r\to \infty$),
we have found that $\Omega^2<0$ in all cases considered.
Thus we conclude that these scalar hairy black hole are unstable against linear
fluctuations.
This result holds also when considering instead the soliton case.

{\bf Static axially symmetric black holes.--}
All known static black hole solutions with scalar hair are spherically symmetric.
However, a Q-ball model possesses also solutions with a spinning phase
\cite{Volkov:2002aj},  
 \cite{Kleihaus:2005me}, 
 \cite{Kleihaus:2007vk},
 \cite{Radu:2008pp}.
 As discussed above, when ignoring the gravity effects, these  Q-balls can be interpreted as
 static, axially symmetric scalarons in a flat spacetime background. 
 The scalar field in this case is complex, with a  phase depending on the azimuthal angle $\varphi$, 
\begin{eqnarray}
\label{s-ans}
\Phi=Z(r,\theta)e^{i n\varphi},
\end{eqnarray}
with $n$ a winding number, $ n=\pm 1,\pm 2,\dots$ (the value $n=0$ corresponds to the
spherically symmetric case discussed above).
These solutions should survive when considering the full model (\ref{action});
therefore we expect to find black hole solutions as well, which are 
static but axially symmetric only\footnote{All known solutions with this property exist 
in models with gravitating non-Abelian fields \cite{Kleihaus:1997ic}.}.

In the numerical construction of such solutions, we have found it convenient to use  
a metric ansatz with three independent functions\footnote{Some of the solutions have been 
recovered by using a Lewis-Papapetrou-type metric Ansatz  instead of (\ref{ansatzg}),
  with 
$
ds^2 =-f dt^2 
+ \frac{m}{f}  \left( {dr^2} + r^2\, d\t^2 \right) 
+ \frac{l}{f}  r^2 \sin^2 \t   d \varphi^2,
$
the functions $f,l,m$ depending on  $r$ and $\theta$ (note that the
radial coordinate here differs from the one used in (\ref{ansatzg})).} 
 \begin{eqnarray}
\label{ansatzg}
ds^2 =- e^{2F_0(r,\theta)}  \Delta(r)  dt^2 
+ e^{2F_1(r,\theta)}  \left( \frac{dr^2}{\Delta(r)} + r^2\, d\t^2 \right) 
  + e^{2F_2(r,\theta)} r^2 \sin^2 \t   d \varphi^2, ~~{\rm where}~~\Delta(r)=1-\frac{r_H}{r},
\end{eqnarray}
and a scalar field given by (\ref{s-ans}).
 $r,\theta$ and $\varphi$ are spherical coordinates; 
 however, the coordinate range for $r$ is $r_H\leq r<\infty$,
with
$r=r_H$ corresponding to an event horizon (thus we do not consider the 
behaviour of the solutions inside the horizon). 
 
A straightforward computation leads to the following expressions for the horizon area,
Hawking temperature and mass of the solutions:
\begin{eqnarray}
\label{quant}
 A_H=4 \pi r_H^2 \frac{1}{2}\int_0^{\pi} d\theta \sin \theta e^{F_1(r_H,\theta)+F_2(r_H,\theta)},
~~
T_H=\frac{1}{4\pi r_H}e^{F_0(r_H,\theta)-F_1(r_H,\theta)},
 ~~
M=\frac{r_H}{2}-c,
\end{eqnarray} 
 where $c$ is a constant which enters the leading order terms in the large $r$ 
expansion of the metric functions,
$
F_1=-\frac{c}{r}+\dots,~~F_2=-\frac{c}{r}+\dots,~~F_0=\frac{c}{r}+\dots.
$

The equations for the metric functions  $ \mathcal F=(F_0,F_1,F_2)$ 
 employed in the numerical calculations,
are found by using a  suitable combination of the Einstein equations  
$E_t^t =0,~E_r^r+E_\theta^\theta =0$ and $E_\varphi^\varphi=0$,
 which diagonalizes them with respect to $\nabla^2 \mathcal F$ 
 (where $\nabla^2=\partial_{rr}+\frac{1}{r}\partial_{r}+\frac{1}{r^2 \Delta}\partial_{\theta\theta}$). 
In the numerical calculations, it turns out to be  convenient to
introduce a new radial variable $\bar r=\sqrt{r^2-r_H^2}$,
such that the event horizon is located at $\bar r=0$.
Then the Einstein-Klein-Gordon equations are solved with the following boundary conditions:
\begin{eqnarray}
\nonumber
\label{bc1} 
\mathcal F\big|_{ \bar r=\infty}=
Z\big|_{\bar r=\infty} =0,~~
\partial_{\bar r} \mathcal F\big|_{\bar r={0}} = 
\partial_{\bar r} Z\big|_{\bar r={0}} =0,
~~
\partial_{\theta} \mathcal F\big|_{\theta={0,\pi}} = 
 Z\big|_{\theta={0,\pi}} =0,
 \end{eqnarray}
 which follow from a study of the 
 approximate form of the solutions, similar to 
 (\ref{sol-hor}),  (\ref{sol-inf}).
The absence of conical singularities
imposes on the symmetry axis the supplementary condition 
 $
 F_1|_{\theta=0,\pi}=F_2|_{\theta=0,\pi},
 $ 
which is  used to verify the accuracy of the solutions. 
Other numerical tests were provided by the Smarr relation (\ref{smarr}) and the first law (\ref{fl}).
Based on that, we estimate a typical relative error around $10^{-3}$ 
for the solutions reported here\footnote{The resulting set of four coupled non-linear partial differential equations is
solved numerically by employing a finite difference solver, 
based on the Newton-Raphson method.
To decrease the errors, a new compactified radial coordinate $x=\bar r/(1+\bar r)$
is introduced.
Then the equations are discretized on a non-equidistant grid in $x$ and $\theta$. 
Typical grids used have sizes $250\times 40$, covering the integration
region $0\leq x \leq 1$ and $0\leq \theta \leq \pi/2$.
}.
Also, all solutions here possess a
reflection symmetry with respect to the equatorial plane, which is used to restrict
the domain of integration for $\theta$ to $[0,\pi/2]$, with 
$\partial_{\theta} \mathcal F\big|_{\theta={ \pi/2}} =\partial_{\theta} Z \big|_{\theta={ \pi/2}}= 0$.

\newpage
 \setlength{\unitlength}{1cm}
\begin{picture}(8,6) 
\put(-0.5,0.0){\epsfig{file=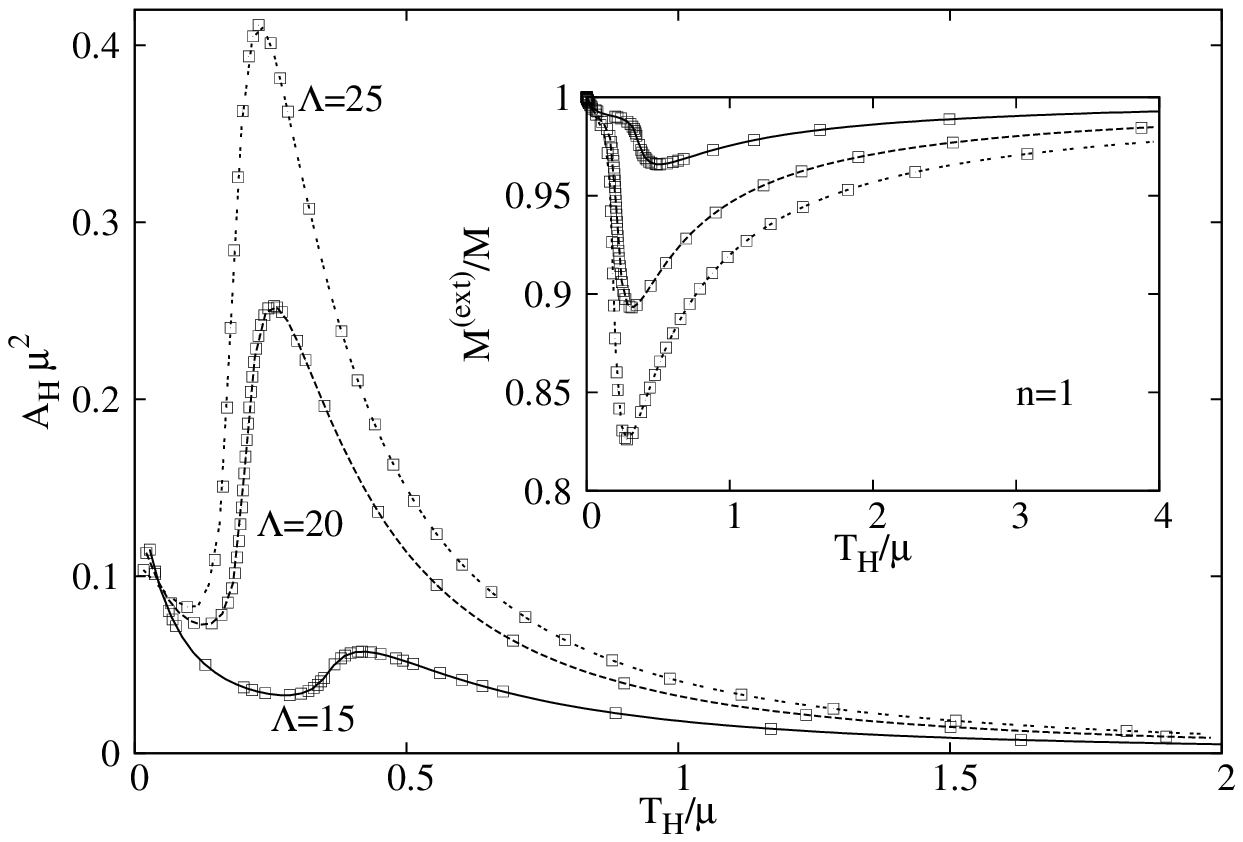,width=8cm}}
\put(8.1,0.0){\epsfig{file=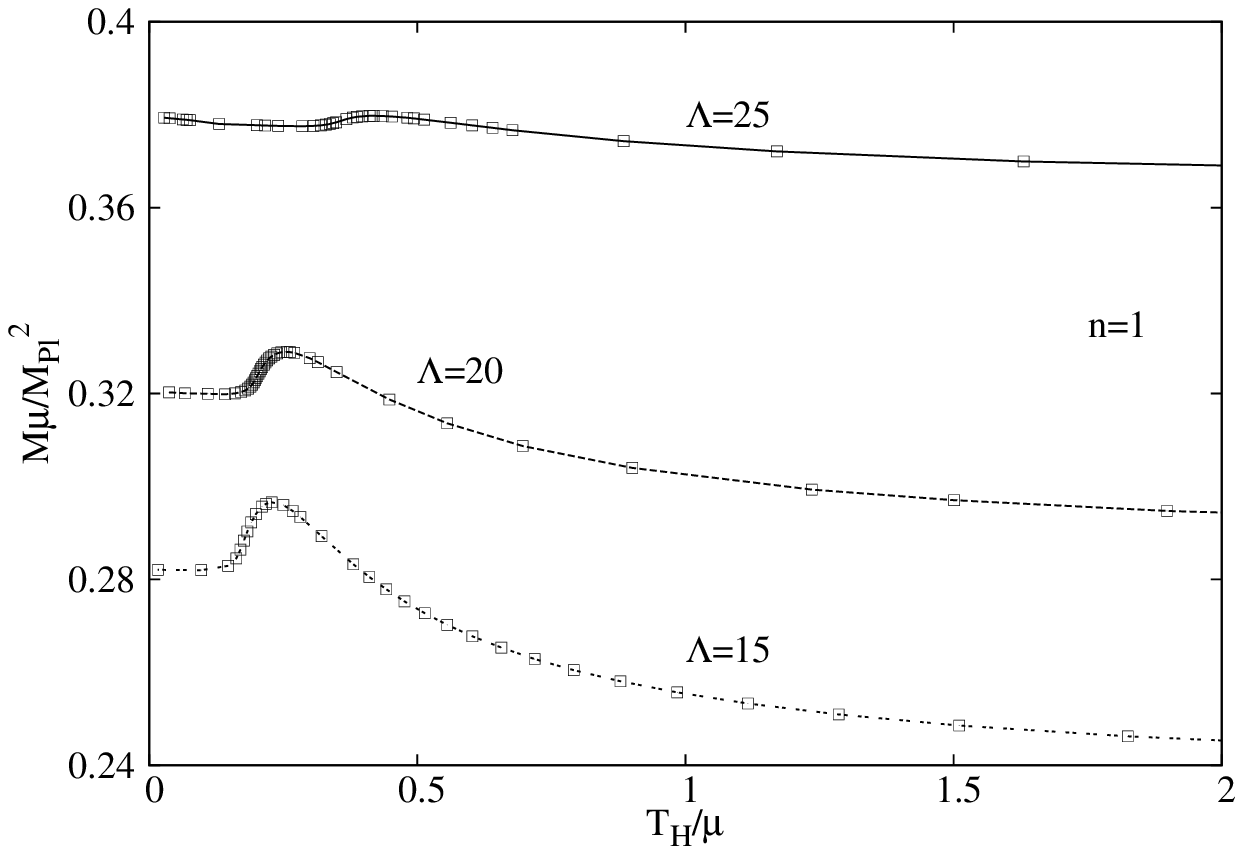,width=8cm}}
\end{picture}
\\
{\small {\bf Figure 3.} 
Some properties of static axially symmetric 
 black holes with scalar hair are shown as functions of the Hawking temperature
for several values of the  dimensionless self-coupling constant $\Lambda$.}
\vspace{0.5cm}
 
The picture in the axially symmetric case is more complicated than
the one found for spherically symmetric solutions.
The solutions reported here have a winding number $n=1$;
however, we have also constructed configurations with $n=2,3$.

As expected, the limit $r_H=0$ still
corresponds to  static solitons 
(the boundary conditions at the origin are
$
\partial_{  r} \mathcal F\big|_{  r={0}} = 
 Z\big|_{ r={0}} =0
 $).
These axially symmetric  scalarons 
have an intrinsic toroidal shape,
as one can see $e.g.$
by plotting surfaces of constant energy density.
Also in this case  $\rho=-T_t^t$ becomes negative in a region close to the origin (however,
for both solitons and black holes, the total mass is still positive).

Again, for any soliton, the origin can be replaced  by a black hole with a small radius.
 However, although it resides at a constant value of $r$,
the horizon is not a round sphere.
This can be seen by evaluating the circumference of the horizon along
the equator, 
$L_e=2 \pi r_H e^{F_2(r_H,\pi/2)}$,
 and the circumference of the horizon along the poles, 
 $L_p=2r_H \int_0^\pi d\theta e^{F_2(r_H,\theta)} $. 
We have found that the ratio $L_e/L_p$ is always slightly
larger than one.

Some thermodynamical
properties of these black hole solutions
are shown in Figure 3
as a function of the temperature (with $T_H\to \infty$ corresponding again to the scalaron limit),
for several values of the self-coupling constant $\Lambda$. 
One can see that the horizon size cannot be arbitrarily large,
while the mass takes smaller values than in the spherically symmetric case 
(note the existence, 
for some range of $T_H$, 
of two different solutions with the same mass).
Another puzzling  feature there is the existence
of a branch of configurations with a positive specific heat.

Also, any family of axially symmetric black holes with a fixed $\Lambda$, 
emerges smoothly from the respective scalaron and
appears to end in a critical configuration.
Extrapolating the numerical results suggests that
the mass and the horizon area
of these critical configurations remain finite,
while the temperature tends to zero.
Constructing such solutions explicitly,
and thus clarifying this limiting behaviour
of axially symmetric black holes, however,
remains a numerical challenge
beyond the purposes of this paper.

{\bf Further remarks.--}
All known asymptotically flat 
scalarons and scalar hairy black holes in the literature
are spherically symmetric and were found
for a rather complicated potential.
In this work we have shown that such solutions exist already 
in a simple model which possesses
only a negative quartic term  in the scalar field potential
in addition to the usual mass term.
Furthermore, black hole solutions with a regular event horizon  which  
are static and  possess only
axial symmetry, do also exist.
Rather unexpected, some of these black holes have a positive specific heat.

In fact, we predict a variety of more complicated solutions to exist.
For example, based on the analogy with 
Q-balls \cite{Kleihaus:2007vk},  odd-parity static axially symmetric scalarons 
and black holes should also exist, with the scalar field vanishing in the equatorial plane.
Moreover, it is likely that the model possesses  
solitons and hairy black holes with discrete crystal-like symmetries only.

Although the analysis in this work was restricted to the (simplest) case of the 
potential (\ref{pot1}),
we expect some basic features of the solutions in this work to be generic.
This conjecture is based mainly on the analogy with  
Q-balls and their gravitating generalizations--boson stars,
which are known to possess a certaing degree 
of universality of the properties, for any potential choice 
(see $e.g.$ \cite{Schunck:1999zu}).
Apart from that, we have verified that the qualitative features of
 the solutions in this work remain unchanged when adding  
 a  positive sextic 
term to the potential (\ref{pot1}), provided that $U$
 is negative for a range of $|\Phi|^2$. 

We hope to return elsewhere with a systematic study of these solutions,
including an existence proof 
for the spherically symmetric case.
Finally, let us  mention
 that
 we have verified that 
the solitons and black holes in this work possess five dimensional
generalizations with rather similar properties.
It is likely that they can be generalized for any spacetime dimension $d>4$.
 
\vspace{0.75cm}
\noindent{{\it \textbf{~~~Acknowledgements.}--~}We gratefully acknowledge support by the DFG,
in particular, also within the DFG Research
Training Group 1620 ''Models of Gravity''. 
\begin{small}

 \end{small}
 
 \end{document}